\documentclass[aps,pra,twocolumn,showpacs,superscriptaddress,10pt,letterpaper]{revtex4-1}

\usepackage[T1]{fontenc}	
\usepackage[latin9]{inputenc}	
\usepackage{geometry}		
\usepackage{graphicx}
\usepackage{amssymb}
\usepackage{hyperref}
\hypersetup{colorlinks=true,urlcolor=blue,citecolor=blue,linkcolor=blue}
\usepackage{listings}

\newcommand{\lrone}{$ACC_1 =  0.566$; $AUC_1 = 0.616$}
\newcommand{\lrtwo}{$ACC_2 = 0.587$; $AUC_2 = 0.614$}

\begin{document}

\title{Examining the relationship between student performance and video interactions}
\author{Robert \surname{Solli}}
\email[Corresponding Author: ]{solli.robert@gmail.com}
\affiliation {Department of Physics \& Center for Computing in Science Education, University of Oslo, N-0316 Oslo, Norway}
\author{John M. \surname{Aiken}}
\affiliation {Department of Physics \& Center for Computing in Science Education, University of Oslo, N-0316 Oslo, Norway}
\author{Rachel \surname{Henderson}}
\affiliation {Department of Physics and Astronomy, Michigan State University, East Lansing, Michigan 48824}
\author{Marcos D. \surname{Caballero}}
\affiliation {Department of Physics and Astronomy, Michigan State University, East Lansing, Michigan 48824}
\affiliation {Department of Physics \& Center for Computing in Science Education, University of Oslo, N-0316 Oslo, Norway}
\affiliation {CREATE for STEM Institute, Michigan State University, East Lansing, Michigan 48824}

\begin{abstract}
In this work, we attempted to predict student performance on a suite of laboratory assessments using students' interactions with associated instructional videos. The students' performance is measured by a graded presentation for each of four laboratory presentations in an introductory mechanics course. Each lab assessment was associated with between one and three videos of instructional content. Using video clickstream data, we define summary features (number of pauses, seeks) and contextual information (fraction of time played, in-semester order).  These features serve as inputs to a logistic regression (LR) model that aims to predict student performance on the laboratory assessments. Our findings show that LR models are unable to predict student performance. Adding contextual information did not change the model performance. We compare our findings to findings from other studies and explore caveats to the null-result such as representation of the features, the possibility of underfitting, and the complexity of the assessment.
\end{abstract}
\maketitle
\section{Introduction}

Recently, video lectures have been implemented as a common instructional tool in many physics classrooms across the country. Most research has focused specifically on the student interaction with the lecture videos. Prior work has shown that students are more likely to finish watching a tutorial laboratory videos in comparison to a lecture videos \cite{lin_gatech_clickstream}. Additionally, students returned to pieces of videos that were specifically related to the laboratory activities \cite{aikenperc2014}. In addition, depending on the content of the video, student ``in-video'' drop out rates have been demonstrated to vary according to the video production quality and the length of the video \cite{lin_gatech_clickstream,guo2014video,kim2014understanding}. Furthermore, students have shown that their decision to watch a specific video depends on how much the video is related to the course content \cite{lin_gatech_clickstream, seaton2014characterizing}.

The implementation of video lectures into physics courses has provided the opportunity for researchers to investigate the effects of viewing and interacting with such instructional materials on student performance. \citet{brinton_pred_perform} found that students' in-video interactions, (e.g., pausing or seeking through the video), were predictive of student performance on the quiz question at the end of the video. While this research is suggestive of a connection between instructional material and student performance, it analyzed video and assessments that are closely connected in time and made use of content assessments that are closely tied to the video content. Our study expands on that result by attempting to predict student performance on tasks that are more complex and separated farther in time from their associated video lectures. 

\citet{lin_gatech_clickstream} investigated student-video interactions in the same course and with the same cohort of students as in our study. In their work, correlations between student-video interactions, average laboratory scores, and performance on exams and FMCE tests were investigated. Given the observed behavior of the student-video interaction with tutorial laboratory videos in this work, we hypothesized that patterns in the student-video interactions would be predictive of their performance on specific laboratory assessments. In our study, we conduct a more granular investigation into whether there was a relationship between the aggregated student-video interaction features (Table \ref{tab:feat}) and the individual grades students earned on the laboratory assessments.

In this paper, we examine the utility of a Logistic Regression (LR) model on predicting whether students received high or low grades on the laboratory assessment. Furthermore, we offer explanations as to why performance on complex activities and assessments might not be predicted from summary student interactions with video lectures. 




\section{Data}\label{sec:data}
Our data was collected in a calculus-based introductory mechanics course in the Fall 2013 semester taught using a flipped model at the Georgia Institute of Technology. At the time of the data collection, the course used the Matter \& Interactions curriculum \cite{chabay_matter} while implementing short video lectures \cite{lin_gatech_clickstream, aikenperc2014} in addition to in-class problem solving and novel laboratory activities \cite{douglas_peer_review, douglas2017peer}. Students' enrolled in the course spent class time focused on various problem solving activities. Outside of the classroom, students spent their time watching lecture videos and performing at-home laboratory activities. 

In total, 161 students were enrolled in the course. The lecture material was delivered in 78 videos throughout the semester. The videos were hosted on the Coursera platform, which logged click events while students were interacting with the videos. 
From these raw clickstream data points, we constructed the features listed in Table \ref{tab:feat}. The features are measures of student interaction with tutorial videos as well as contextual information about the student and video in question. Each row-entry of this matrix we denote as a feature vector. The laboratory-videos retained a high percentage of viewer-ship throughout the semester as compared to the lecture-themed videos, which saw a time-dependent decay of access rates \cite{lin_gatech_clickstream}. Because students continued to watch laboratory tutorial videos, we were able to directly model student performance using feature vectors on multiple laboratory assessments.

Throughout the semester, the students performed four laboratory assessments -- each coupled with one to three lecture videos.
For each laboratory assessment, students were required to make a presentation as a way to demonstrate their understanding of the covered material. The students made their presentation over a two week period \cite{lin_gatech_clickstream}. The presentations were then peer-evaluated and graded \cite{douglas_peer_review}. These grades have been shown to serve as a reasonable proxy for expert grades \cite{lin_peer_review, douglas_peer_review, douglas2017peer}.
Students were allowed to drop one of the laboratory grades and therefore, some students did not participate in all exercises; these entries were graded as zero and were removed from the analysis. 

In addition to the laboratory presentations, students' conceptual understanding was measured by the Force and Motion Conceptual Evaluation (FMCE) \cite{fmce} both pre- and post-instruction.
Students' demographic information was not available.

\begin{table}[t]
\caption{The features used in modeling student performance on laboratory assessments. The first 5 features (with asterisks) were used in one model with the additional 3 features used to a subsequent model. 
}\label{tab:feat}
\begin{tabular}{l|p{60mm}}
\hline Feature name & Description \\
\hline \hline
$\mathrm{t_N}$ * & Time each student spent watching a video, normalized by the video length, averaged - Z-scored  \\ \hline
$\mathrm{t_A}$ * & Number of individual accesses to each video  averaged Z-scored. \\ \hline
pauses * & Number of pauses with automatic pauses removed in the video averaged - Z-scored. \\ \hline
plays * & Number of plays averaged - Z-scored.\\ \hline
seeks * & Number of seeks averaged - Z-scored. \\  \hline
interaction time & Average time relative to the population mean of the student interaction with videos, averaged - Z-scored \\ \hline
lab & Which lab the feature vector corresponded to in the course sequence \\ \hline
$\mathrm{FMCE_{pre}}$ & Each students score on the FMCE test prior to taking the course.  \\ \hline
\hline
\end{tabular}

\end{table}

\section{Methods}\label{sec:methods}
In this work, we attempted to predict whether the student received a high or low grade (split on the median) on the laboratory presentation using the model features presented in Table \ref{tab:feat}. 
As we have designed this study as a binary classification problem, we have chosen to use a supervised learning approach in lieu of linear regression.
Below, we describe how we implement a Logistic Regression (LR) model \cite{hosmer2013applied} for this study.

\paragraph{Preprocessing}\label{p:process} Prior to analyzing any model, preprocessing the data is important. As there was only one grade for each set of lecture videos (Sec.~\ref{sec:data}), we constructed our model data such that each graded presentation for a laboratory exercise held one average entry for each student. For example, for a student that clicked the pause button twice in the first video and four times in the second, the pause feature for the associated laboratory exercise for the student would hold the value of $3$. There were then $N = 627$ data points in the model data. Additionally, many of the features varied largely in magnitude and were not normally distributed. Regularized Logistic Regression (LR) models require the features to have a mean of zero mean and unit variance, thus the individual features were scaled \cite{mendenhall2016statistics}.

\paragraph{Training the models}\label{p:train} 
Because a supervised learning approach was taken to explore the effect of video lectures on student performance, the data was partitioned into a training set and a validation set. For this analysis, the fraction of data in the training set was chosen to be 60\%, which is common for classification tasks \cite{olson2017data}. Finding the optimal model using LR require adjusting hyper-parameters (i.e., the regularization strength, $C$; the stopping criterion, $tol$; and the maximum iterations, $\textrm{max\_iter}$) to find the model that best predicts the outcome without overfitting. Tuning the hyper-parameters is commonly done by sweeping through a large range of values at random (a random search) to find a reasonable starting point and, then by conducting a more targeted sweep of nearby values (a grid search).
For the random search, the regularization strength took on random values between $\{0 < C \leq 100\}$, the stopping criterion took on values between $\{0 < tol \leq 100\}$, and the maximum number of iterations ranged from $\{0 < \textrm{max\_iter} \leq 200\}$. The best combinations of randomized hyper-parameters,  as judged by their prediction quality, were subjected to a grid search to further tune the hyper-parameters. 
Step sizes in the grid search were 10\% of the value of each hyper-parameter, and the grid search took 5 steps in both directions. The best-performing model, in both search cases, was determined using the measures presented below. 
\paragraph{Estimating model performance}\label{sec:est}
Once the model has been trained, the model performance is then evaluated on the validation set via the accuracy score (ACC) and the area under the curve for the Receiver Operating Characteristic (ROC). The ACC of a model is simply defined as the percentage of correctly classified entries in the dataset. The Receiver Operating Characteristic (ROC) is a well established measure of a classifiers performance \cite{swets_roc}. The performance is measured as the effect of the true positive rate (TPR) and the false positive rate (FPR) as a function of thresholding the positive class (i.e. high performing students). To evaluate the ROC curve for a model, one traditionally uses the Area Under the Curve (AUC) \cite{bradley_auc} which ranges from $0$ (a perfect ``opposite'' classifier)  to $1.0$ (a perfect classifier) with $0.5$ indicating a random guess classifier.
Testing the null-hypothesis that the classifier does no better than random guessing is done by the Kolmogorov-Smirnov (KS) test \cite{bradley_roc}. To estimate if the LR model was underfitting the data a Random Forest (RF) \cite{RF_Breiman2001} and a Support Vector Machine (SVM) \cite{vapnik2013nature} model was trained on the same data, and subjected to the same metrics. 

\section{Results}\label{sec:results}

\begin{figure}
\includegraphics[width = 0.85\columnwidth, height = 2.2in]{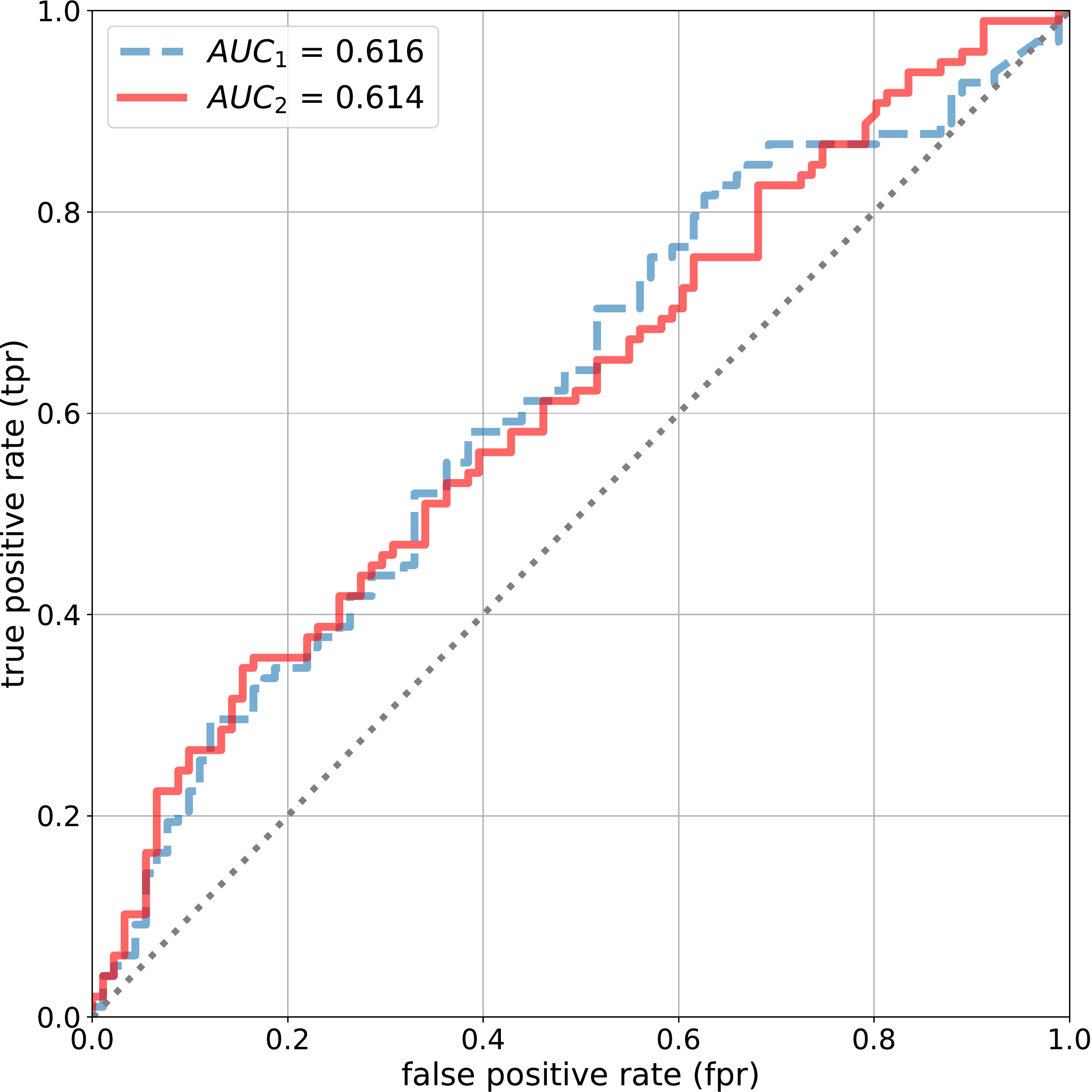}
\caption{Plots of the ROC for both LR Model 1 (dashed) and LR Model 2 (solid) on the same training-test partitioning of the data. The x-axis represents the fraction of samples in the test set falsely classified (FPR) as positive (high performers), and the y-axis gives the true fraction of the positive (TPR). The dotted diagonal line represents the performance of the trivial random guess model.}\label{fig:roc_plots}
%
%
\end{figure}

LR models were trained with the procedure outlined in Sec.~\ref{sec:methods}. For LR Model 1, only the first 5 features in Table \ref{tab:feat}, which focus entirely on how student interact with the videos, were used in the model. This model is represented by the dashed non-diagonal line in Fig.~\ref{fig:roc_plots}. LR Model 1 performed marginally better than a model uniformly guessing at the dependent variable, specifically the student's grade on the laboratory assessment as being above or below the population median (\lrone).

As LR Model 1 was a poor predictor of student performance solely from student-video interactions, some contextual features from the course, including when the videos appeared in the course and an initial measure of student conceptual understanding, were added to the model (LR Model 2). The average interaction time and the course sequence feature were included to control for the sensitivity to the context of the viewing behavior.
In addition, the student's FMCE pretest score was included to control for the varying degrees of incoming conceptual understanding. The ROC curves for LR Model 2 are also included in Fig.~\ref{fig:roc_plots} as the solid line (\lrtwo).

Neither the student-video interaction model (LR Model 1) nor the second, enhanced model with contextual features (LR Model 2) were able to successfully predict student performance on laboratory assessments in this context. To validate whether these models performed better than chance, a Kolmogorov-Smirnov (KS) goodness-of-fit test between each ROC and the distribution for chance prediction (i.e., the diagonal in Fig. \ref{fig:roc_plots}) was computed: LR Model 1 ($D_n$ =  $0.084$  $p =0.772$) and LR Model 2 ($D_n = 0.118$  $ p = 0.335$). Neither model was statistically significant and therefore, both models were not predictive of student performance on the laboratory assessment.

\section{Discussion \& Conclusion}

The original hypothesis was that there was a relationship between the aggregated click-stream features and a student's grades on laboratory presentations. 
Our analysis indicates that predicting student performance from summary clickstream quantities was not possible for these laboratory presentations. Although \citet{brinton_pred_perform} presented an AUC $> 0.7$ for models predicting correct answers on an end-of-video quiz from similar summary clickstream data, the complexity of the laboratory assessments could be impacting our ability to successfully predict student outcomes. In the \citet{brinton_pred_perform} study, clickstream data was used to predict student performance on an assessment that was close in time to the video interaction. Furthermore, the assessment used by \citet{brinton_pred_perform} was relatively simple by comparison (i.e., recalling and/or applying knowledge from the video). In this study, the laboratory assessment activity was arguably more complex, requiring that students not only recall and apply their knowledge, but also to engage with that knowledge through scientific practices such as developing models, constructing explanations, and communicating scientific information. So while these tutorial videos might provide some initial information needed to engage in the laboratory assessment, they do not appear to be a major factor in determining student performance. When course contextual information such as FMCE pretest scores were included (LR Model 2), it did not improve classification performance. 

That being said, there might be a number of reasons why the methodology that we employed might give pause to our null result.
There is the possibility that the models are underfit because the number of data points were low ($N = 627$). We also cannot discount the effect of our representation of the model features as direct measurements instead of the probabilistic estimates utilized by \citet{brinton_pred_perform}. Furthermore, a number of our features followed Poisson distribution and thus normalizing those features using z-score might introduce marginal issues.  Additionally the effect of representing the dependent variable as a binary value split at the median of the score might be ill suited to the problem; however, this was chosen due to ill fitting linear regression models ($R^2=0.056$). Further work could include exploration of a different number of classes.

While there are a number of possible challenges to our methodology, we do have some confidence that we are indeed finding no connection between patterns of student-video interaction and laboratory assessments. We investigated the performance of RF and SVM models to understand if the scaling of features or low complexity in the LR model impacted the predictive performance. In head-to-head comparisons, RF and SVM tend to out perform LR in a variety of classification tasks \cite{olson2017data}. Neither model was able to produce a classifier with a significant KS statistic when compared to the random guess classifier ($D_n < 0.125$ $p > 0.28$). The RF model was constructed using unscaled features and as such provides evidence that the effect of breaking the assumption of normality when z scaling the features was negligible. 


While {\it post facto}, our results might seem obvious, that is, interacting with a tutorial video in different ways is not suggestive of different levels of performance on laboratory activities in our context, we believe that this analysis is critical to understanding what manner of course materials support our students engaging in the more complex tasks. This is especially true with the growth in flipped models of instruction where students watch videos of different types at home in order to engage with more complex reasoning, modeling, and/or experimental activities in class.

In our particular case, how students use tutorial laboratory videos does not appear to be important for predicting their performance on more complex laboratory activities. So while these videos might be serving the purpose providing some initial information about the laboratory activities, how students watch them is not indicative of how they perform in the lab. From a curriculum and instruction perspective, this suggests that videos in a flipped course might not serve the precise purpose that the designer intended and that more intentional and innovative designs might be needed to be explored. For a research perspective, our results suggest that a good model of student performance on complex activities, like the laboratory assessments in this course, likely starts from other forms of engagement in the course, for example, those activities which designers intend for students to engage with in-class.

\acknowledgments{This study was funded by the Olav Thon Foundation and the Norwegian Agency for Quality Assurance in Education (NOKUT), which supports the Centre for Computing in Science Education.}
%

%
%

\bibliography{bibliography}
\end{document}